# Social Similarity-aware TCP with Collision Avoidance in Ad-hoc Social Networks

Hannan Bin Liaqat, Feng Xia, *Senior Member, IEEE*, Jianhua Ma, Laurence Tianruo Yang, Ahmedin Mohammed Ahmed, *Student Member, IEEE,* and Nana Yaw Asabere

*Abstract*—Ad-hoc Social Network (ASNET), which explores social connectivity between users of mobile devices, is becoming one of the most important forms of today's internet. In this context, maximum bandwidth utilization of intermediate nodes in resource scarce environments is one of the challenging tasks. Traditional Transport Control Protocol (TCP) uses the round trip time mechanism for sharing bandwidth resources between users. However, it does not explore socially-aware properties between nodes and cannot differentiate effectively between various types of packet losses in wireless networks. In this paper, a socially-aware congestion avoidance protocol, namely TIBIAS, which takes advantage of similarity matching social properties among intermediate nodes, is proposed to improve the resource efficiency of ASNETs. TIBIAS performs efficient data transfer over TCP. During the course of bandwidth resource allocation, it gives high priority for maximally matched interest similarity between different TCP connections on ASNET links. TIBIAS does not require any modification at lower layers or on receiver nodes. Experimental results show that TIBIAS performs better as compared against existing protocols, in terms of link utilization, unnecessary reduction of the congestion window, throughput and retransmission ratio.

## I. Introduction

ADVANCES in wireless communication technologies have reinforced the development and use of Internet of Things (IoT). IoT can be regarded as a global network infrastructure that employees current evolving Internet and links physical and virtual objects. This link is achieved through exploring communication capabilities of various networks. Ad-hoc Social Network (ASNET) is a branch of IoT networks that are based on the social properties or behaviors of users. In most cases, transmission of application data exchange and social metadata updates create congestion in ASNETs. In application data exchange, congestion is high due to multimedia data, file

- *This work was partially supported by the Natural Science Foundation of China under Grant No. 60903153, Liaoning Provincial Natural Science Foundation of China under Grant No. 201202032, and the Fundamental Research Funds for the Central Universities.*
- *H. B. Liaqat, F. Xia, A. M. Ahmed, and N. Y. Asabere are with School of Software, Dalian University of Technology, Dalian 116620, China.*
- *J. Ma is with Faculty of Computer and Information Sciences, Hosei University, Japan.*
- *L. T. Yang is with Department of Computer Science, St. Francis Xavier University, Canada.*
- *Corresponding author: F. Xia; Email: f.xia@ieee.org*

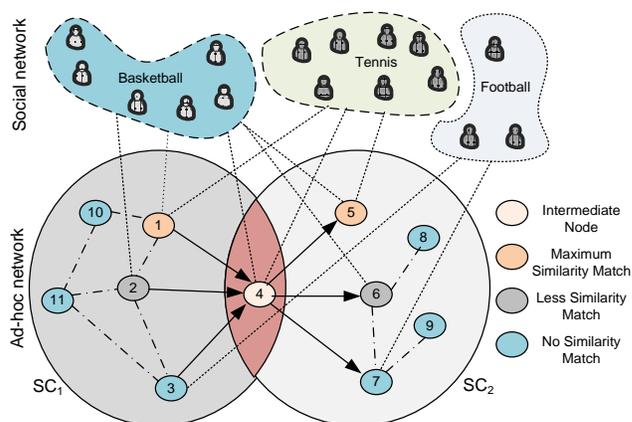

Fig. 1.  Social similarity based ASNET model

transferring and multiple users communicating at simultaneous times. In distributed environments, every node shares its own profile with neighbor nodes and tries to find those ad-hoc nodes that have similar interest, otherwise the nodes serve as a relay [1][2]. Therefore, social metadata creates overhead in ASNETs in addition to the exchange of application data. Short handshake messages are used for sharing the social profile of each node, which recognizes the communicating nodes. Optionally, the nodes also exchange full social profiles if there has been a change since the last contact between the nodes. After a successful handshake, social networks execute the forwarding algorithm for matching the discovered node's social profile to the data messages carried by the node.

Nodes in ASNETs are connected directly or through a multi-hop based on some social properties, such as social graph, community, centrality, similarity, tie strength and human mobility pattern [3][4]. Fig. 1 explains the ASNET model, which uses the social property of similarity for communicating and finding the nodes of similar interest in the two communities; a key feature of real social network as noted by Buscarino *et al.* [5]. The model consists of two layers, the virtual social network layer and the physical ad-hoc network layer. In a virtual social network layer, nodes with common interest are connected by logical link. The communication between distant nodes in the physical ad-hoc network layer is according to message transmission of multiple hops. Therefore, one hop logical link in the mobile social network may be composed of several hops in the ad-hoc network layer. Each node manages the communication of the nodes that have common interest.

Let us explain the rationale behind Fig. 1 through the following example. We categorize nodes in the network into two Social Community $SC_1$ and $SC_2$. The nodes in $SC_1$ (1, 2 and 3) are sender nodes that want to communicate with receiver nodes. Receiver nodes belongs to $SC_2$ which are 5, 6 and 7. Nodes in the sender and receiver sides have some social interests that relate to the social network layer. In Fig. 1 there are three social interests shown in the social network layer, consisting of Basketball, Tennis and Football. The single physical node that lies in ad-hoc network layer may have one or more social interests. Due to this interest matching situation, nodes provide maximum advantage or high priority for the same interest. Therefore, we exploit similar interests based matching for data transfer. Due to the longer distance between sender and receiver nodes, the $SC_1$ sender nodes will use intermediate node 4 for communication. We categorize a node's social properties into three types, namely maximum similarity match, less similarity match and no similarity match. A sender node is referred as maximum similarity match if it has the same social interests with the intermediate node. If a sender node matches some social interests with an intermediate node, it is referred as less similarity match. On the other side, if there is no social interest matched with an intermediate node, then this is called no similarity match. Generally, the intermediate node receives multiple data from the sender nodes. However, it cannot serve all requests, as there may be insufficient bandwidth, which often leads to the first congestion limitation. Moreover, congestion loss can also occur due to the social property of the 'similarities' of intermediate node. An intermediate node prefers not to share maximum bandwidth with those sender nodes that have less or no social similarities. On the other hand, nodes that have strong matched social similarities with an intermediate node can transfer at a high data rate. Traditional Transport Control Protocol (TCP) [6] cannot perform expeditiously in such social interest-matching environments, because the setting of traditional TCP data rate is based on Round Trip Time (RTT). Therefore, we intend to tackle this limitation by adjusting the data rate to match the social properties between nodes.

The second limitation in ASNETs is packet loss in terms of wireless link error. Traditionally, whenever the loss occurs, TCP assumes such loss is due to congestion. However, the packet loss in ASNETs may not always be due to congestion. In an attempt to correct this packet loss, traditional TCP reduces the size of the congestion window, and further minimizes the throughput of the whole network. This results in severe setbacks if such an assumption was incorrectly perceived. As such existing packet loss differentiation techniques are not suitable for ASNETs. Existing strategies differentiate between congestion and wireless losses through traditional TCP congestion window, explicit congestion notification and RTT policies. However, ASNET should differentiate packet loss using congestion window based on the social similarity property without breaking end-to-end semantic for effective communication. In order to avoid congestion related loss and random loss, none of the existing approaches provide a proper solution simultaneously, especially in ASNETs. The congestion in ASNETs occurs due to the socially inherent limitations of intermediate node and multiple users using multiple applications. In addition, these packet losses affect the performance of the network, because congestion related losses increase unnecessary retransmission and wireless related losses increase unnecessary reduction of the congestion window.

Our proposed solution in this paper, namely TIBIAS, avoids the above mentioned limitations by providing an efficient solution to improve the performance of TCP. Our targets are to reduce the congestion related losses and avoid unnecessary reduction of the window at transport layer. The transport layer can provide more reliability with end-to-end connectivity and handle congestion related losses accurately, since data rate is effectively set on this layer. The main contributions of this paper are: Firstly, a socially-aware approach, i.e. TIBIAS, is proposed for avoiding congestion in mobile IoT or mobile ad hoc networks. Secondly, TIBIAS is used to differentiate between various types of random packet losses and it reacts accordingly. Our proposed scheme is a sender side mechanism that circumvents the congestion loss related to scarce bandwidth using social property (i.e. similarity) of nodes. In contrast to existing proposals, TIBIAS avoids congestion related losses by exploiting the social interest of intermediate nodes and provides high data rate for maximum matched social similarity. Also, it provides the least data rate to those nodes which have less or no social similarity with the intermediate node. Furthermore, the proposed protocol differentiates between random packet losses to avoid the unnecessary reduction of the congestion window, consequently enhancing the network Quality of Service (QoS) in ASNETs.

The rest of this paper is organized as follows. Section II provides a brief review of related literature. Section III presents an overall detailed structure of our TIBIAS solution. Section IV describes the socially-aware congestion avoidance and differentiation module in detail. Section V explains the bandwidth estimation module and Section VI discusses the similarity-based bandwidth allocation module of TIBIAS. In Section VII, extensive simulation results are presented and analyzed, followed by conclusion in Section VIII.

## II. RELATED WORK

In ASNETs nodes interconnect each other directly or through intermediate ones. ASNET communications are often based on social properties, such as similarity, centrality, social relationship and community. Rahnama *et al.* [7] emphasized that "Adaptation of Social Context" is an important feature in the design of social networking systems. However, to implement this, there is a need for a generic algorithm that can adapt itself to changing social context of ad hoc networks. Moreover, ASNETs suffer from many types of losses: 1) multiple applications run on a single connection and hence create congestion, 2) intermediate nodes drop packets due to less social similarity match between nodes, and 3) wireless link related error. Besides the scarcity in bandwidth, intermittent connectivity and high bit-error-rate also create problems for ASNETs [8]. The remaining part of this section discusses the emergence of ASNETs and the enhancement of TCP in wireless networks.

*A. Ad-hoc Social Networks*

ASNET is a new emerging area and an interconnection between social networks and mobile IoT, where nodes meet on the basis of social characteristics and store social profile or records in local mobile devices. Recently, some applications that facilitate the mobile nodes to search common interest nodes are available. For instance, E-SmallTalker [9] provides a mechanism to find common interest topics through Bloom Filter without establishing a connection. MobiClique [2] uses profile-based matching, by providing synchronization between profiles through Facebook API, and connects to a profile server. Terry *et al*. [10] provided users with the facilitation for interest matching using collocation pattern. Moreover, some researchers are working to design relevant systems that provide connectivity between nodes using social properties of ad-hoc nodes, e.g., in [11] and [12]. To facilitate object discovery, Liu *et al*. [13] presented socially-aware like systems that depend on self-managing P2P topology with human tactics for social networks. For scale-free P2P networks, SP2PS [14] provides a solution for resource discovery with low maintenance overhead cost. Furthermore, regarding resource discovery in structured and unstructured P2P systems, there are quite a lot of related research efforts (e.g. [15] and [16]). Finding an appropriate route between nodes is a key challenge for data packet in ASNETs. Therefore, integrating social behaviors into such networks makes it easier to establish an appropriate route. For example, the authors of [17] proposed to use the social property of nodes to establish a best route. Similarly, the authors in [18]-[20] used social properties of nodes to forward the data and reduce the congestion using social properties. However, exploiting social properties to set data rate over TCP has not yet been explored in the context of ASNETs.

*B. TCP-based Wireless Networking*

The improvement of TCP for wireless networks has been a crucial research point in the past few years. A number of researchers have made efforts on addressing packet losses using various techniques, such as Random Early Detection [21], Explicit Congestion Notification (ECN) [22], Explicit Transport Error Notification [23] and Explicit Loss Notification (ELN) [24].

Some new techniques like TCP-CERL [25] and Supervised Learning [26] only distinguish between the types of packet losses. TCP-CERL measures RTT for estimating the queue length of the router and finds out the reason for losses. Further research works in relation to TCP performance in wireless networks have been done in [27] and [28] by classifying causes of packet losses. Conversely, TCP Peach [29], TCP Peach Plus [30], and TCP Westwood [31] make different modifications in congestion control after bandwidth estimation. Meanwhile, TCP Jersey [32] differentiates between packet losses and handles congestion control but the drawback of this protocol is that it breaks the concept of TCP end-to-end semantics. On the other hand, TCP Veno [33] provides the solution between congestion and random losses. It provides sender side solution and maintains end-to-end TCP semantic but in an environment such as a lightly loaded wireless network, Veno suffers from severe bandwidth under utilization [34]. Another protocol, ESTCP [35], provides distinction between losses and avoids congestion related loss using dynamic AIMD congestion window. Similar to TCP Jersey, ESTCP also uses ECN bit for controlling the congestion related loss and breaks the concept of TCP semantics.

Despite existing related works with different characteristics, none of them has defined the allocation of data rate based on social properties of nodes. Carofiglio *et al*. [36] and Rozhnova *et al*. [37] defined the interest rate control in wired networks. Amadeo *et al*. [38] also defined the interest rate control in a Content Centric Network (CCN) based on wireless ad-hoc networks. However, the strategy of these protocols for controlling data rate is not based on social properties of nodes. The allocation of data rate in our scheme is different from CCN and these techniques are implemented in content centric networks, not for ASNETs. In our approach, we will improve the performance of existing TCP, using social properties of nodes to allocate the data rate in ASNETs.

## III. STRUCTURE OF TIBIAS

The two objectives of TIBIAS are to attain the full available bandwidth consumption and to fulfill the requirement of user similarities in terms of sharing the congested bandwidth among multiple ASNET sender nodes. The detailed description of this process is demonstrated in Section IV. There is no assumption of prior knowledge of the network in TIBIAS, and thus predicting the available capacity in advance is too complex in the case of ASNETs.

In traditional TCP, the total available bandwidth is consumed by integrating throughput of the flow. The objective of our system is to utilize the capacity of a network fully as traditional TCP; but the data rate is based on interest similarities of the user's. The fundamental responsibilities of the proposed system are limited to provide bandwidth to low priority users or those users who have limited match interests with the intermediate node. TIBIAS further assigns extra bandwidth to higher interest matched users.

TIBIAS does not provide fairness in TCP connections at the sender side but uses an intermediate node that constrains the sender's congestion window. In Fig. 2, we define our proposed system block diagram. The Socially-aware Congestion Avoidance and Differentiation Module (SCADM) is the first section of our TIBIAS system, where we can get the desired data rate of a TCP connection after matching user's interest similarities and available bandwidth. The input value in SCADM is Similarity Interest Rate ($SIR_m$) and Bandwidth Estimation ($BE_m$); and the desired interest Congestion Window (*cwnd*) rate $C_m$ is the output of SCADM. SCADM has two sub-parts to set the value of the desired congestion window, Socially-aware Congestion Avoidance Sub-module (SCAS) and Socially-aware Differentiation Sub-module (SDS). SCAS provides the actual available $BE_m$ after calculating the difference between actual estimated rate and desired $SIR_m$. Moreover, SDS provides a differentiation module between losses and reacts accordingly after receiving N duplicate acknowledgment (N-dupack). In Section V, we provide details about the bandwidth estimation process of our system. SCADM provides an easy mechanism to utilize the maximum link capacity and to calculate the desired data rate for different nodes in ASNETs. Although this mechanism achieves the intermediate based similarity-matching goal, every TCP connection in the network may not utilize full link capacity because some flows in the network falls short and due to this reason, the sender will set the data rate under the desired level. TIBIAS provides a solution for efficient utilization of available bandwidth by overcoming the above-mentioned challenges.

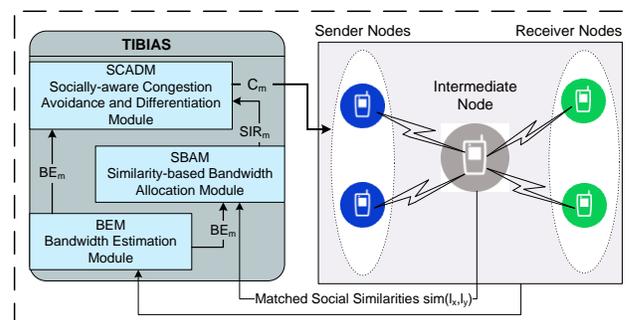

Fig. 2. TIBIAS model for social-aware communications in ASNETs

TIBIAS operates with *m* TCP connections and actual measured rate will be an input to the Bandwidth Estimation Module (BEM), which shows the estimated value of a network. $BE_m$ is the output of BEM and represents the available bandwidth of a single TCP connection in the presence of multiple TCP connections. $BE_m$ is responsible for providing desired data rate in SCADM and Similarity-based Bandwidth Allocation Module (SBAM); it is also responsible for the actual available capacity of the specific TCP connections. Fig. 2 illustrates the estimated value of bandwidth $BE_m$ and the matched social similarity $Sim(l_x,l_y)$ of an intermediate node. The SBAM will be described in Section VI and provides the desired $SIR_m$ to SCADM after estimating $BE_m$ and matched social similarities between nodes. In the next section, we delve into the details of these TIBIAS components.

## IV. SOCIALLY-AWARE CONGESTION AVOIDANCE AND DIFFERENTIATION MODULE

SCADM provides two sub-modules as depicted in Fig. 3, one relates to Socially-aware Congestion Avoidance Sub-module (SCAS), while the other is related to Socially-aware Differentiation Sub-module (SDS). In SCAS, after receiving an acknowledgement, the specific data rate is assigned based on matched social similarity of an intermediate node. However, when the sender node receives N-dupack, SDS differentiates between losses and then manages $C_m$ based on similarity rate control. SCADM controls the losses related to congestion and sets the window value proactively.

As shown in Fig. 4, the flowchart of SCADM provides a detailed overview of TIBIAS system. It adopts the idea of slow-start and fast recovery from TCP Reno [39]; slow-start is helpful for estimating the bandwidth. In a slow-start phase, the congestion window is increasing exponentially. After reaching the Slow Start Threshold (*ssthresh*), socially-aware congestion avoidance phase starts where increment and decrement in congestion window is based on matched interest similarity. Socially-aware congestion avoidance phase is different from TCP Reno because the movement of the congestion window is not linear. We adopt the concept related to explicit retransmission from TCP Jersey [32], but our protocol replaces rate control procedure in [32] and instead utilizes similarity rate control. In explicit retransmission, Jersey does not divide the congestion window into halves and retransmits the loss data explicitly with a recent congestion window. Similar to TCP-Jersey, the adjustment of the congestion window is based on the similarity rate control after checking the reason of loss. Similarity-rate control procedure is explained in the next subsection.

$SIR_m$ and $BE_m$ are the inputs of the SCADM and the output of the system is $C_m$ that achieves the desired data rate. SCADM takes input $SIR_m$ as a value; estimates the bandwidth of the entire network to find out the network capacity for a specific TCP connection. SCADM continuously upgrades the congestion window to achieve the desired interest matched data rate. In the subsequent sections below, we define the steps to achieve the desired match data rate.

### A. Socially-aware Congestion Avoidance Sub-module

After a slow start phase, congestion avoidance phase begins, which controls congestion rate. Our module provides the last desired interest rate *cwnd* $C_l$ after using the fraction £ of the similarity interest rate $SIR_m$, and then measures the $BE_m$. The purpose of our system is to calculate the value of congestion window $C_l$ within a fraction £ of similarity interest rate $SIR_m$ and sets updated desired interest rate *cwnd* $C_m$ accordingly i.e. $C_l \in ((1-£)SIR_m,(1+£)SIR_m)$. In our system, the fraction value £=0.3 and we assume that the congestion window provides packets of equal segment size with an integer value; *Seg_size* and the minimum value of $C_m$ must be equal or greater than 1. In the slow start phase, congestion window moves exponentially after reaching *ssthresh*, and then socially-aware congestion avoidance phase also starts. Here *Seg_size* is the size of sent segment packet and RTT is calculated by average round trip time of the flow. To calculate the RTT value, we use the TCP timestamp option in [40]. In our system, we assume that the congestion window $C_a$ uses actual available bandwidth rate, round trip time and per segment size.

$$C_a = BE_m * RTT / Seg\_size \quad (1)$$

$C_b$ needs to set the value of a window according to the similarity interest rate. The value of $C_b$ is:

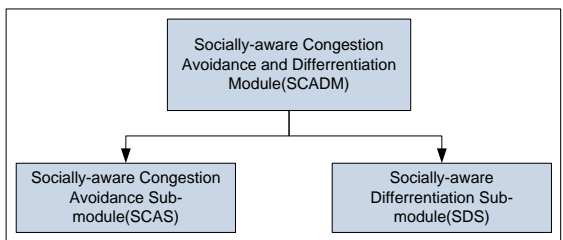

Fig. 3. Sub-modules of SCADM

$$C_b = SIR_m * RTT/Seg\_size \quad (2)$$

Estimated values of $BE_m$ and $SIR_m$ provide the accurate value of social similarities to set the value of $\Delta C$, which depends on the $\Delta D$ by differentiating (1) and (2):

$$\Delta C/\Delta D = RTT/Seg\_size \quad (3)$$

Before defining the Socially-aware Congestion Avoidance Sub-module (SCAS) algorithm, we define the strategy of the system to avoid congestion. Most of the time, it is observed that SCAS enters in a state quickly where $C_l < SIR_m$. In this situation, the value of the congestion window needs to be increased, however; $C_l$ reaches the desired interest rate. The following steps define how SCAS works to avoid congestion related loss.

Initially, if the congestion window $C_l$ lies between the fractional value of £ with desired similarity interest rate $SIR_m$ i.e. $C_l \in ((1-£)SIR_m, (1+£)SIR_m)$, there is no need to make a change. Hence, assign $C_m = C_l$.

Secondly, if $C_l > (1+£)SIR_m$, then $C_l$ decreases for achieving the desired interest rate $C_m$. The value of $C_m$ should be greater than or equal to 1 when no social similarity matches. In order to get a decrease in $C_m$, we use (3) as follows.

$$\Delta C = \{(\Delta D)(\frac{RTT}{Seg\_size}) + \gamma\} \quad (4)$$
$$C_m = C_l - \Delta C \quad (5)$$

Equation (5) operates repetitively until the $C_m$ is less than or equal to $SIR_m$. Here we define $C_l$ as the previous congestion window and the value of $\gamma$ is equal to 0.5 due to linear decrease. $\Delta C$ provides the difference between social similarities $SIR_m$, original $BE_m$ with respect to $RTT$ per $seg\_size$. After that SCAS subtracts the $\Delta C$ from $C_l$ and gets the desired $C_m$.

In the final step, if $C_l < (1-£)SIR_m$, then we need to increase $C_m$ using (3):

$$\Delta C = \{((\Delta D)\left(\frac{RTT}{Seg\_size}\right))*\gamma\} \quad (6)$$
$$C_m = C_l + \max(\Delta C, 1) \quad (7)$$

For adjusting the factor of the congestion window, we multiply $\Delta C$ by $\gamma$; the limit of the adjustment factor must be less than 1 in a single iteration and we adjust the value of $\gamma$ to be equal to 0.8. The value of $C_m$ should be increased by 1, because we want to leave the state where congestion window is less than the desired similarity interest rate. The change in congestion window shows changes in throughput of the system. Bandwidth estimation and round trip time create influence on the value of the congestion window. Therefore, it is important to estimate bandwidth accurately for these two parameters.

After SCAS, the system checks the value of Retransmission Time-Out (RTO); if RTO expires, it moves into the slow-start phase, otherwise it waits for acknowledgment. If it receives N duplicate acknowledgment (N-dupack), the system will move it to SDS, which is described in the next section, otherwise it moves again to the SCAS.

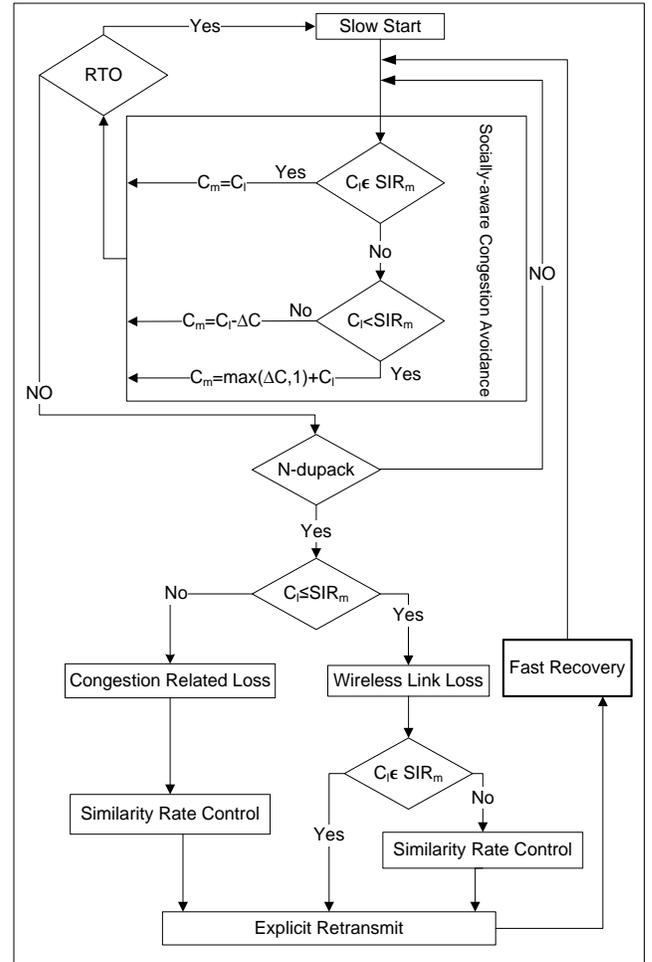

Fig 4. Flow diagram of SCADM

### B. Socially-aware Differentiation Sub-module

The reason of packet loss in ASNET is not only related to congestion. Wireless related loss can also be the reason of packet loss as shown in Fig. 4. To identify the reason of loss after calculating the social properties of the system, TIBIAS has a mechanism that provides the sender side with a solution. To identify the cause of loss, existing parameters include RTO, RTT and N-dupack. In this regard, TIBIAS uses the N-dupack parameter for differentiation.

TIBIAS detects the reason of loss using the condition between $C_l$ and $SIR_m$ after receiving N-dupack. If $C_l \leq SIR_m$, it shows that the loss is because of a wireless link error or random loss. Next to detecting the wireless related loss, there is a need to further sort out the conditions. If $C_l \in SIR_m$, the loss of packets are retransmitted explicitly without addition or reduction in the congestion window. However if $C_l < SIR_m$ then it is moved in similarity rate control, where it needs to modify the data rate to achieve the desired $C_m$ using (7). Meanwhile if $C_l > SIR_m$, it shows the loss related to congestion. Therefore it moves to similarity rate control and needs to set desired $C_m$ using (5). After assigning the updated value to $C_m$, the similarity rate control will transfer to explicit retransmission and then move to the recovery phase. After recovery phase, the data will move into SCAS phase to avoid the congestion.

## V. BANDWIDTH ESTIMATION

Estimating the Bandwidth in an ad-hoc environment is a complicated process because every time the environment changes along with its wireless conditions. Therefore, bandwidth estimation techniques need special attention to estimate the bandwidth value. In ASNETs, bandwidth is also an intricate challenge, because lots of bandwidth is consumed due to the multiple applications running simultaneously on user side. Therefore, bandwidth estimation is a difficult issue for ASNETs. After estimating the available bandwidth, BEM provides estimated value to SBAM and SCADM. In SBAM, the purpose of using this bandwidth estimation value is to assign the specific available bandwidth after matching the social similarities between nodes, while SCADM uses this actual bandwidth estimation value $BE_m$ to subtract from $SIR_m$ for achieving the $C_m$.

In this section, we define the lightweight bandwidth estimation algorithm that does not require a lot of acknowledgment history to estimate the bandwidth. We denote that the overall bandwidth of the system as $BE_m$ and to calculate this value we elaborate on the following procedure.

$$SR_m = C_l / \text{RTT}_{min} \quad (8)$$

$SR_m$ defines the minimum sending rate of the data using the previous available $C_l$ congestion window with minimum Round Trip Time ($RTT_{min}$). Equation (8) shows the data rate being sent within one RTT interval.

To calculate the value of the sample bandwidth estimation, $BE_s$, the $ACK\_Data$ is calculated in bytes and divided by time $t_m$. Equation (9) shows the sample bandwidth results. To calculate the value of $t_m$, in (10) our BEM uses the $SR_m$ for sending rate and subtracts it from $BE_{m-1}$. After subtraction, the difference value is multiplied by RTT and divided by $SR_m$.

$$BE_s = Ack\_Data / t_m \quad (9)$$
$$t_m = (SR_m - BE_{m-1}) * RTT / SR_m \quad (10)$$

To calculate the overall bandwidth estimation, BEM is defined in the following procedure.

$$\sigma = \frac{4C_l - Ack\_Data}{4C_l + Ack\_Data} \quad (11)$$
$$BE_m = \sigma BE_{m-1} + (1-\sigma) BE_S \quad (12)$$

In (11), $\sigma$ shows that the coefficient of variation at time $t_m$, coefficient variation can be achieved after the subtraction of the last congestion window from acknowledgment data and addition of last congestion window from acknowledgment data. In (12), $BE_m$ shows the overall bandwidth estimation of the network using the coefficient variation $\sigma$; $BE_{m-1}$ is the previous value of the estimated bandwidth and a sample rate of bandwidth estimation $BE_s$.

## VI. SIMILARITY BASED BANDWIDTH ALLOCATION

In Section IV, the TCP connection at the sender nodes achieves the desired throughput after matching the $SIR_m$ and $BE_m$. Now, we consider the scenario where multiple senders share the same intermediate node. The aim of our proposed system is to allocate bandwidth among multiple TCP connections using an intermediate interest similarity-matching scheme. Our next goal is to provide the full link utilization of intermediate node after matching the similarities. $SIR_m$ is the output of this module and $BE_m$, matched social similarities $sim(l_x,l_y)$ is the input of this module. The assignment of $SIR_m$ is based on the available bandwidth and social similarities matched between sender and intermediate node. The detailed methodology is defined in the remaining parts of this section.

### A. SBAM Operation

The flow of connections in TCP depends on the social applications that are only applicable when a particular bandwidth threshold is set between users having the same interest. Meanwhile, if no interest similarity is matched between users or low similarities are matched between users, TCP needs to set the minimum bandwidth required for the minimal data rate. These limitations are required for assigning the allocated bandwidth to each TCP connection. After assigning a minimum bandwidth, TCP needs to assign the remaining bandwidth according to matched similarities. The sender node sets bandwidth assignment rate from lower to higher priority.

In Section IV, we have shown that SCADM provides the desired interest data rate for a single TCP connection. SCAS achieves throughput by setting the sender side window using the $SIR_m$. We already mentioned in TIBIAS overview that SBAM uses the value of $BE_m$ and the matched social similarities to provide $SIR_m$ to SCADM.

$BE_m$ is already measured in Section V to provide the similarity based data rate allocation for the entire system using $m$ as multiple connections, sharing intermediate node and $SIR_m$ is the similarity interest desired rate for the $m^{th}$ connection. Once the bandwidth estimation module sets $BE_m$, SBAM uses $BE_m$ to derive the desired data rate $SIR_m$ per TCP connection.

### B. Similarity and Bandwidth Allocation Measurement

In this subsection, we explain the logic of how to partition the interest rate. To calculate the interest similarity $Sim$, our system uses similarity formula from [1]. To compute the similarity between two nodes profiles', we consider $l_x$ and $l_y$.

$$Sim(l_x, l_y) = \frac{\sum_{u=1}^{u=z} \max_{k \in \{1,f\}} Sim(dx_h, dy_k)}{z} \quad (13)$$

In (13) $z$ shows the number of concepts in profile $l_x$ and the number of concepts in $l_y$ is denoted by $f$. $Sim(l_x,l_y)$ compares with user defined similarity threshold ($0 < threshld \le 1$). If $Sim(l_x,l_y) > threshld$ then $l_x$ is semantically related with $l_y$, if $Sim(l_x,l_y) = threshld$ it shows that $l_x$ is equally related to $l_y$ and if $Sim(l_x,l_y) < threshld$ it shows that $l_x$ has less similarity with $l_y$. In our scenario, the value of social similarities must be greater than 0 and the value of $threshld$ is 0.6 as we defined in our simulations.

To calculate $SIR_m$, we consider $C_l$ that is defined in multiple segments to explain the occupancy of ASNETs. The $SIR_m$

function mentioned below is taken into account and corresponds to the maximum similarity based data rate at connection *m* controlling the transmission window under *RTT*, where *k* represents the design parameter, which manages the dynamic nature of $SIR_m$ to achieve $BE_m$. Generally, the available bandwidth is distributed among all active connections when *T* connections connect through an intermediate ad-hoc node. Each communication in our scheme shares and matches the social similarity part of $BE_m/T$ of the total bandwidth. Using a value less than $BE_m/T$ causes the remainder resources to be allocated window sized fluctuations based on RTT. If $C_l$ is the number of packets from last conversation then the $SIR_m$ for communication *m* is represented as:

$$SIR_m = Sim(l_x, l_y) + k * \frac{BE_m/T - C_l}{RTT_m} \quad (14)$$

The value of $Sim_{int}(l_x, l_y)$, limits the similarity interest rate and represents the maximum capacity of sending the similarities at the $m^{th}$ connection. $SIR_m$ is shown below

$$SIR_m = min(max(SIR_m, 0), Sim_{int}(l_x, l_y)) \quad (15)$$

Equation (14) shows that similarity can possibly become negative. Under this condition, the allocation priority of similarities is in descending order, up to the minimal rate of each TCP connection. Equation (15) illustrates that each TCP connection will get the minimum data rate, according to its matched similarity rate. Maximum matched similarity will assign a high data rate and the remaining bandwidth will be assigned according to its social priority. This section will give the social similarity rate and compares the available $C_l$ to *SIR*. The next section discusses our detailed results of simulation and comparison with existing TCP variants.

## VII. Performance Evaluation

The proposed method, TIBIAS, is evaluated through the OPNET simulation tool [41] using metrics such as throughput, link utilization, and retransmission number of packets. It is also compared against other existing variants of TCP in ad-hoc social environments.

### A. Simulation Setup

The system model in Fig. 5 consists of existing approaches such as ESTCP [35], TCP-CERL [25], TCP Reno [39], TCP Veno [33] and TIBIAS. Our main objective is to investigate whether TIBIAS can perform on link utilization, unnecessary reduction of the congestion window, throughput and number of retransmissions, when compared with other protocols. We consider a similarity based bandwidth allocation module for controlling the congestion and differentiating between losses. SBAM contains several similar interests that indicate interest based communication and intermediate node forward data based on matched similarities. It utilizes semantic based user profile matching that uses queries to find the similar profile between two ASNET nodes. Initially five sender and receiver nodes are deployed with one intermediate node.

The common parameters in TIBIAS simulation are demonstrated in Table I. For the implementation and comparison, we used 802.11b ASNET and categorized it into a small medium network size, traffic load and wireless link error conditions.

The network size comprises of five and twenty nodes for small and medium size network respectively. Each network includes objects such as ad-hoc nodes, ad-hoc gateway, definition of application objects and profile information. Ad-hoc nodes use ad-hoc gateway to make connections with IP and relay the FTP traffic from source to destination. Whereas ad-hoc network gateway relates to a wireless local area network that has a single Ethernet interface. All the ad-hoc nodes and gateway have the same DSR routing protocol configuration. DSR is a reactive routing protocol that helps to reduce the routing cost and overhead of the network.

In order to define different applications we use the application object. The applications may be video conferencing and FTP. These applications are used in user profiles to offer the similarities or similar interests between ad-hoc users.

We modified the basic Reno with the slow-start initial value equal to 1 MSS and used Jersey explicit retransmission method. For matching the interest similarities, different types of interest are set in the profile configuration system. We analyzed the multiple TCP variants using packet loss rate with link utilization (Fig. 8(a)) and unnecessary reduction of congestion window (Fig. 8(b)), the effect of number of connections on throughput (Fig. 9(a)) and retransmission number of packets (Fig. 9(b)), and then we compared throughput based on propagation delay (Fig. 10(a)), bandwidth (Fig. 10(b)) and queue size (Fig. 10(c)). The concept in [42] is adopted to evaluate the throughput. Additionally we analyze the performance of TCP variants in the total number of retransmission packets with packet loss rate and multiple TCP connections.

### B. Social Metadata Management Cost

Comparison between two user profiles on the basis of common ontology graph can be the tool to pre-compute the pair wise distances in order to obtain the result easily for future computation. It is worth noting that similarity is not a symmetric relation which means that similarity between A and B might give a different result from similarity between B and A. To evaluate the cost of social metadata, initially we measure the size of the profile attributes with delay and then secondly we measure network management overhead with the number of connections. We used four similarity threshold, 0.2, 0.4, 0.6 and 0.8 to evaluate the ontological similarity.

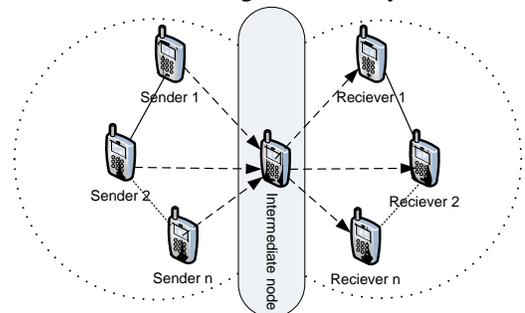

Fig. 5 Simulation setup for multiple ad-hoc nodes

In the scalability of the profile matching algorithm, we measure the time cost to compute the similarity between profiles with increased number of properties. The data is collected from a population of users of our ASNET model. The number of interests possessed by each user was varied and then required latency was measured to conduct comparison between profiles of different sizes. Fig. 6 demonstrates that as a number of property increases, the latency does not show much change, which proves a profile matching algorithm scalable in terms of the profile size.

In this paper, the mean of overhead relates to the number of times a social metadata has to be exchanged for the management of the network when new connections occur. The system finds interested nodes and starts matching functionality. The maximum interest matching function provides the extra bandwidth to the sender nodes. Likewise, the average delay will increase if all profile attributes are used in matching functions. Therefore, it is interesting to understand how many times overhead cost take place per efficient management of a social metadata, and which threshold is the most efficient. Fig. 7 depicts the management overhead per threshold and the number of connections. The most efficient threshold strategy and the reference case is that of *threshld* 0.6. In this strategy, matching attributes are not too much and create average overhead on the network. Here we can see that a social metadata management overhead cost is less than $3.0*10^{5}$ up to $4^{th}$ node. The management overhead measures the update times of the social metadata within the given period. As compared to the exact-matching, multifold has more relevant friends that were identified by ontology-based similarity matching technique. The reason for using this similarity identification function is to reduce the social metadata overhead cost and manage the network easily. This concept uses semantic level rather than the syntax level to measure the similarity [1]. Therefore, two profiles can be semantically related despite being literally different. However, an increase in the number of the results can also be observed when the similarity threshold decreases, which causes a network overhead. With this true relationship, the number of connections can also increase the network overhead and create difficulty to manage the network. Therefore, TIBIAS users construct an ASNET with an average

TABLE I
SIMULATION PARAMETERS

| Simulation Parameters | Values | Units |
|---|---|---|
| No. of TCP connections | 1-20 | - |
| Area | 250*255 | *m\*m* |
| Data rate | 6 | *Mbps* |
| Transmission power | 0.007 | *Watts* |
| Ad-hoc node buffer | 50 | *KB* |
| Data type | FTP | - |
| Maximum delay ack | 0.2 | *s* |
| Duplicate ack threshold | 3 | - |
| Duration of simulation | 2000 | *s* |
| No. of simulation | 10 | - |
| Message buffer size | 50 | *Packets* |
| Slow-start initial count | 1 | *MSS* |
| Max RTO | 64 | *s* |
| Min RTO | 0.5 | *s* |

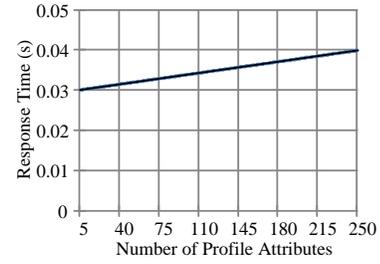

Fig. 6. Impact of profile attributes size on latency

threshold value 0.6 to reduce the network management and overhead cost.

C. *Results and Analysis*

We consider the performance of multiple variants in the presence of congestion loss and random loss, and further examine the performance according to multiple metrics for the evaluation of TIBIAS. Fig. 8(a) demonstrates the trend line between percentage of link utilization and packet loss rate. The wireless ad-hoc link between nodes has a propagation delay of 50 *ms* with a bandwidth of 6 Mbps. The range of packet loss is between $10^{-7}$ and $10^{-2}$. From Fig. 8(a), it can be seen that the increase in packet loss rate leads to decrease in the percentage of link utilization. Additionally, our proposed scheme outperforms in general the other approaches by achieving better link utilization.

Fig. 8(b) demonstrates the relationship between unnecessary reduction of congestion window and the rate of packet loss. The rate of packet loss varies from 5% to 10%. In general, the increase in packet loss also increases the unnecessary reduction of the congestion window packets. TCP Reno reduction rate is higher than all other variants because in Reno, the consideration of every loss is due to congestion. The unnecessary reduction limit in TIBIAS is almost similar with ESTCP when loss rate increases but the performance of TIBIAS is better than ESTCP and all other variants. TIBIAS's unnecessary reduction of packet is 14 at 10% loss rate; however, other variants, such as ESTCP, TCP CERL, TCP Veno and TCP Reno have 17, 19, 20 and 23 unnecessary reduction of packets respectively.

Fig. 9(a) illustrates the throughput of TCP with multiple TCP connections. The results show that by increasing the number of connections, data throughput is decreased. The TCP connections range from 1 to 20 with 3% random loss, 26 Mbps bandwidth, and 50 *ms* delay rate. TCP CERL has stability in throughput after 17 TCP connections and achieves 21 Mbps throughput. Our proposed scheme has an overall better throughput in multiple TCP connections whereas existing approaches show drastic changes in throughput. The performance of TIBIAS is 14.5% higher than ESTCP and 94% higher than TCP Reno.

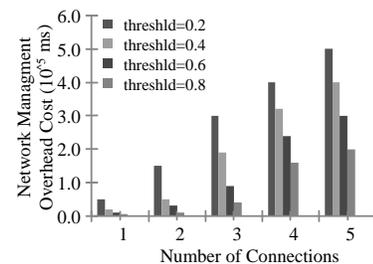

Fig. 7. Network management overhead vs. number of connections while measuring similarity with different threshold ranges

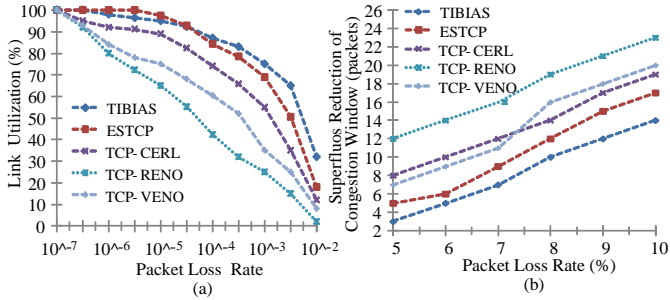

Fig. 8. Impact of packet loss rate on (a) link utilization and (b) superfluos reduction of congestion window

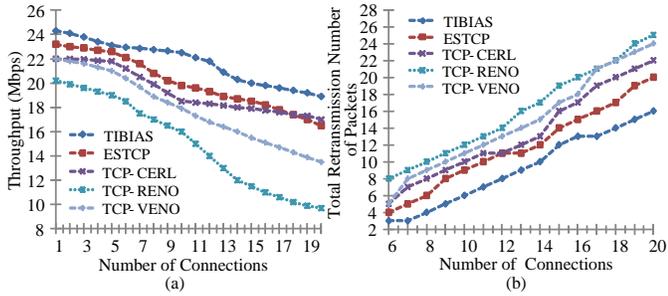

Fig. 9. Impact of number of connections on (a) throughput and (b) total retransmission number of packets

Fig. 9(b) illustrates the dependency between TCP connections and retransmission number of packets. The TCP connections vary from 6 to 20 and the range of retransmission depends on the number of TCP connections. Multiple connections to the same intermediate node increase the number of retransmission packets because the chance of loss also increases. Our protocol performs better than all other existing TCP variants and executes lower retransmission. The performance of TIBIAS is higher due to its proactive congestion control methodology and assigns data rates based on social similarities between the nodes after estimating the bandwidth. TIBIAS retransmits a total 16 packets when the number of TCP connections is 20.

Fig. 10(a) demonstrates the performance of multiple TCP variants in terms of data throughput with respect to propagation delay of links. The wireless ad-hoc link between nodes has a loss rate of 2% with a bandwidth of 6 Mbps. The propagation delays vary from 50 to 150 (*ms*). It is evident from Fig. 10(a) that the increase in delay decreases the size of congestion window, which reduces throughput. Overall, our proposed scheme performs more efficiently; the throughput is increased when compared to the other existing approaches. TIBIAS utilizes its delay advantage after verifying the accurate reasons for loss. When TIBIAS receives N-dupack, it determines the desired interest rate value. If *cwnd* is less than the desired rate, it specifies different categories of loss consisting of random loss or wireless related loss. After detecting the reason of random loss, congestion window needs to be increased, in case *cwnd* has less value than the desired rate. Our proposed scheme provides maximum throughput because it offers 28% improvement over TCP Reno, 25% improvement over TCP Veno, and delivers 16.3% improvement over ESTCP at propagation delay of 150 *ms*.

Fig. 10(b) shows throughput with respect to bandwidth. The packet loss rate is 3% and the delay between ad-hoc nodes is 50 *ms*. The bandwidth changes from 10 to 40 Mbps. The increase in throughput is noticed by increasing the bandwidth in our conducted analysis. The performance of the proposed scheme is better than existing schemes. TCP Reno and TCP Veno do not perform efficiently as utilization of bandwidth is less than all other protocols.

Fig. 10(c) depicts the change of throughput with the size of queue. The size of queue is generally proportional to throughput. As the size of queue increases, it also intensifies the rate of throughput. The size of queue ranges from 40 KB to 80 KB. TIBIAS covers this utilization of queue after estimating the bandwidth of the network, setting the rate of data according to available bandwidth, and matching desired interest rate. The performance of TIBIAS as compared to ESTCP is 5.3% higher and 23% higher than TCP Reno. TCP Veno and CERL both achieve 14.3 Mbps at 80 KB size of queue.

## VIII. CONCLUSION

This paper has proposed an innovative resource allocation technique (called TIBIAS) for ASNETs, where nodes communicate with each other based on socially-aware properties. TIBIAS is totally based on transport layer and provides reliability with end-to-end connectivity; controls congestion related losses and detects the reason for loss after receiving the N-dupack. TIBIAS controls the congestion based on socially-aware properties after estimating the bandwidth of ASNET nodes. Using socially-aware properties, TIBIAS adopts the similarity property where there is an interest-based match between nodes, and then based on matched interest results, TIBIAS sets the data rate for congestion control.

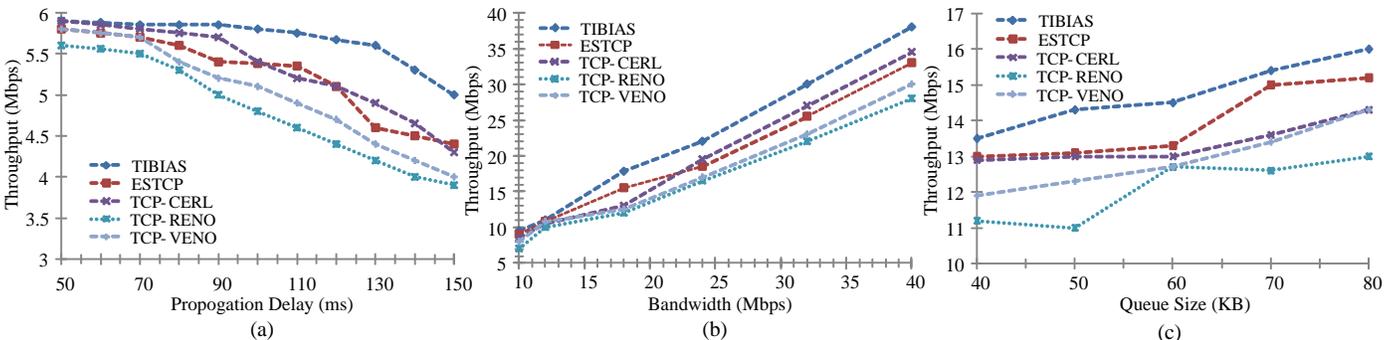

Fig. 10. Throughput with respect to (a) propagation delay, (b) bandwidth and (c) queue size

TIBIAS can easily detect the losses from ASNETs and provides simple sender side solution. We compare TIBIAS' performance with existing TCP variants such as ESTCP, TCP-CERL, TCP Reno and TCP-Veno. The simulation results show that TIBIAS avoids unnecessary reduction in the congestion window after receiving the N-dupack. Hence, TIBIAS provides maximum throughput and efficient utilization of link in high error rate. In a nutshell, TIBIAS yields better performance as compared against these existing TCP variants.

As future works, we will extend TIBIAS to different cases of social relationships in multiple communities. Many design issues need to be considered such as multiple hop data transferring, scalability and fairness. Better concession should be found among these issues to take full advantage of them. We also envisage the possibility of making a new protocol that manages data packet based on social properties (similarity). In such data management protocol design, we may utilize state of the art data filtering and dropping techniques.